\documentclass[american,aps,prl,reprint,superscriptaddress,showpacs]{revtex4-1}
\usepackage[T1]{fontenc}
\usepackage[latin9]{inputenc}
\usepackage{color}
\usepackage{babel}
\usepackage{amsthm}
\usepackage{amsmath}
\usepackage{amssymb}
\usepackage[unicode=true,pdfusetitle,
 bookmarks=true,bookmarksnumbered=false,bookmarksopen=false,
 breaklinks=false,pdfborder={0 0 0},backref=false,colorlinks=true]
 {hyperref}
\hypersetup{
 linkcolor=urlcolor,citecolor=urlcolor,urlcolor=urlcolor}

\makeatletter
\@ifundefined{textcolor}{}
{%
 \definecolor{BLACK}{gray}{0}
 \definecolor{WHITE}{gray}{1}
 \definecolor{RED}{rgb}{1,0,0}
 \definecolor{GREEN}{rgb}{0,1,0}
 \definecolor{BLUE}{rgb}{0,0,1}
 \definecolor{CYAN}{cmyk}{1,0,0,0}
 \definecolor{MAGENTA}{cmyk}{0,1,0,0}
 \definecolor{YELLOW}{cmyk}{0,0,1,0}
 }
\theoremstyle{plain}
\newtheorem{thm}{\protect\theoremname}
\theoremstyle{plain}
\newtheorem{lem}[thm]{\protect\lemmaname}
\ifx\proof\undefined
\newenvironment{proof}[1][\protect\proofname]{\par
\normalfont\topsep6\p@\@plus6\p@\relax
\trivlist
\itemindent\parindent
\item[\hskip\labelsep
\scshape
#1]\ignorespaces
}{%
\endtrivlist\@endpefalse
}
\providecommand{\proofname}{Proof}
\fi

\usepackage{babel}
\usepackage{babel}
\usepackage{babel}
\usepackage{babel}
\usepackage{babel}
\usepackage{babel}
\usepackage{babel}
\usepackage{babel}
\usepackage{babel}
\usepackage{babel}
\usepackage{txfonts}\usepackage{braket}

\definecolor{urlcolor}{rgb}{0,0,0.7}

\definecolor{eric}{rgb}{0,.5,.2}

\newcommand{\mbb}{\mathbb}
\newcommand{\mc}{\mathcal}

\newcommand{\op}[2]{|#1\rangle \langle #2|}

\providecommand{\theoremname}{Theorem}

\providecommand{\lemmaname}{Lemma}
\providecommand{\theoremname}{Theorem}

\providecommand{\lemmaname}{Lemma}

\providecommand{\theoremname}{Theorem}

\providecommand{\lemmaname}{Lemma}

\providecommand{\theoremname}{Theorem}

\providecommand{\lemmaname}{Lemma}
\providecommand{\theoremname}{Theorem}

\providecommand{\lemmaname}{Lemma}
\providecommand{\theoremname}{Theorem}

\providecommand{\lemmaname}{Lemma}
\providecommand{\theoremname}{Theorem}

\providecommand{\lemmaname}{Lemma}
\providecommand{\theoremname}{Theorem}

\providecommand{\lemmaname}{Lemma}
\providecommand{\theoremname}{Theorem}

\providecommand{\lemmaname}{Lemma}
\providecommand{\theoremname}{Theorem}

\makeatother

\providecommand{\lemmaname}{Lemma}
\providecommand{\theoremname}{Theorem}

\begin{document}

\title{Assisted distillation of quantum coherence}

\author{E. Chitambar}

\affiliation{Department of Physics and Astronomy, Southern Illinois University,
Carbondale, Illinois 62901, USA}

\author{A. Streltsov}

\email{streltsov.physics@gmail.com}

\author{S. Rana}

\author{M. N. Bera}

\affiliation{ICFO -- Institut de Ciències Fotòniques, Av. C.F. Gauss, 3, E-08860
Castelldefels, Spain}

\author{G. Adesso}

\affiliation{$\mbox{School of Mathematical Sciences, The University of Nottingham, University Park, Nottingham NG7 2RD, United Kingdom}$}

\author{M. Lewenstein}

\affiliation{ICFO -- Institut de Ciències Fotòniques, Av. C.F. Gauss, 3, E-08860
Castelldefels, Spain}

\affiliation{ICREA -- Institució Catalana de Recerca i Estudis Avançats, Lluis
Companys 23, E-08010 Barcelona, Spain}

\date{January 15, 2016}
\begin{abstract}
We introduce and study the task of assisted coherence distillation.
This task arises naturally in bipartite systems where both parties
work together to generate the maximal possible coherence on one of
the subsystems. Only incoherent operations are allowed on the target
system while general local quantum operations are permitted on the
other, an operational paradigm that we call local quantum-incoherent
operations and classical communication (LQICC). We show that the asymptotic
rate of assisted coherence distillation for pure states is equal to
the coherence of assistance, an analog of the entanglement of assistance,
whose properties we characterize. Our findings imply a novel interpretation
of the von Neumann entropy: it quantifies the maximum amount of extra
quantum coherence a system can gain when receiving assistance from
a collaborative party. Our results are generalized to coherence localization
in a multipartite setting and possible applications are discussed.
\end{abstract}


\maketitle
\noindent \textbf{\emph{Introduction}}\textbf{.} Quantum coherence
represents a basic feature of quantum systems that is not present
in the classical world. Recently, researchers have begun developing
a resource-theoretic framework for understanding quantum coherence
\cite{Aaberg2006,Levi-2014a,Baumgratz2014,Bromley2015,Streltsov2015,Yuan2015,Korzekwa2015,Winter2015,Singh2015}.
In this setting, coherence is regarded as a precious resource that
cannot be generated or increased under a restricted class of operations
known as incoherent operations \cite{Levi-2014a,Baumgratz2014}. A
resource-theoretic treatment of coherence is physically motivated,
in part, by certain processes in biology \cite{Lloyd-2011a,Li-2012a,Huelga-2013a},
transport theory \cite{Rebentrost-2009a,Bjorn-2013a,Levi-2014a},
and thermodynamics \cite{Lostaglio2015a,Lostaglio2015b,Korzekwa2015},
for which the presence of quantum coherence plays an important role.

In this paper, we consider the task of \emph{assisted coherence distillation}.
It involves (at least) two parties, Alice ($A$) and Bob ($B$), who
share one or many copies of some bipartite state $\rho^{AB}$. Their
goal is to maximize the quantum coherence of Bob's system by Alice
performing arbitrary quantum operations on her subsystem, while Bob
is restricted to just incoherent operations on his. The duo is further
allowed to communicate classically with one another. Overall, we refer
to the allowed set of operations in this protocol as \textbf{\emph{L}}\emph{ocal
}\textbf{\emph{Q}}\emph{uantum-}\textbf{\emph{I}}\emph{ncoherent operations
and }\textbf{\emph{C}}\emph{lassical }\textbf{\emph{C}}\emph{ommunication}
\emph{(LQICC)}. As we will show, the operational LQICC setting reveals
fundamental properties about the quantum coherence accessible to Bob.
In particular, the von Neumann entropy of his state, $S(\rho^{B})$,
quantifies precisely how much extra coherence can be generated in
Bob's subsystem using LQICC than when no communication is allowed
between him and any correlated party.

Alice and Bob's objective here is analogous to the task of assisted
\textit{entanglement} distillation. In the latter, entanglement is
shared between three parties, $A,B,C$, and the goal is for $B$ and
$C$ to obtain maximal bipartite entanglement when all parties use
(unrestricted) Local Operations and Classical Communication (LOCC).
The corresponding maximal entanglement that can be generated between
$B$ and $C$ is known as ``entanglement of collaboration'' \cite{Gour2006}.
Henceforth, here we define the ``coherence of collaboration'' as
the maximum coherence that can be generated on subsystem $B$ by LQICC
operations. In general, both LOCC and LQICC protocols can be very
complicated, involving many multiple rounds of measurement and communication
\cite{Chitambar2014}. A simplified scenario considers one-way protocols
in which Alice holds a purifying system, and only she is allowed to
broadcast measurement data. The maximum entanglement for $B$ and
$C$ (resp.~maximum coherence for $B$) that can be generated in
this manner is called the ``entanglement of assistance'' \cite{DiVincenzo1999}
(resp.~will be called the ``coherence of assistance''). In the
asymptotic setting the entanglement of assistance is known to be equal
to the entanglement of collaboration if the overall state is pure
\cite{Smolin2005}. We show an analogous result for coherence: for
pure states the coherence of assistance is equal to the coherence
of collaboration in the asymptotic setting, and a closed expression
for these quantities is also provided.
Moreover, when Bob's system
is a qubit and the overall state is pure, the coherence of assistance
and the coherence of collaboration are equivalent even in the single-copy
case.
Finally, we also present a generalization to a multipartite
setting where many assisting players collaborate to localize coherence
onto a target system, and discuss possible applications to quantum
technologies.

\noindent \textbf{\emph{Resource theory of coherence}}\textbf{.} The
starting point of our work is the resource theory of coherence, introduced
recently in \cite{Levi-2014a,Baumgratz2014,Bromley2015,Winter2015}.
In particular, a quantum state $\rho$ is said to be incoherent in
a given reference basis $\{\ket{i}\}$, if the state is diagonal in
this basis, i.e., if $\rho=\sum_{i}p_{i}\ket{i}\bra{i}$ with some
probabilities $p_{i}$. For a bipartite system, the reference basis
is assumed to be a tensor product of local bases \cite{Bromley2015,Streltsov2015,Winter2015}.

A quantum operation is said to be incoherent, if each of its Kraus
operators $K_{\alpha}$ is incoherent, i.e., if $K_{\alpha}{\cal I}K_{\alpha}^{\dagger}\subseteq{\cal I}$,
where ${\cal I}$ is the set of incoherent states.  In this theory, a general completely positive trace-preserving (CPTP) map $\Lambda$ is said to be incoherent if it can be represented by at least one set of incoherent Kraus operators.  Completely dephasing
any state $\rho$ in the incoherent basis will generate the incoherent
state $\Delta(\rho):=\sum_{i}q_{i}\op{i}{i}$ with $q_{i}=\braket{i|\rho|i}$.  Note this is entire motivation for defining incoherent states as being diagonal in some particular basis: they are the density matrices obtained by erasing all off-diagonal terms through the decoherence map $\Delta$.  If $d$ is the dimension of the Hilbert space of the system, the maximally
coherent state is $\ket{\Phi_{d}}=\sqrt{1/d}\sum_{i}\ket{i}$, and
we let $\ket{\Phi}:=\ket{\Phi_{2}}$ denote the ``unit'' coherence
resource state \cite{Baumgratz2014}.

Similar to the framework of entanglement distillation \cite{Bennett1996a,Bennett1996},
general quantum states can be used for asymptotic distillation of
maximally coherent states via incoherent operations.  Formally, the distillable coherence $C_d$ of a state $\rho$ is defined as $C_d(\rho)=\sup\left\{ R:\lim_{n\rightarrow\infty}\left(\inf_{\Lambda}\left\Vert \Lambda\left[\rho^{\otimes n}\right]-\Phi^{\otimes\left\lfloor Rn\right\rfloor }\right\Vert \right)=0\right\}$, where $\left\Vert M\right\Vert=\mathrm{Tr}\sqrt{M^\dagger M}$ is the trace norm, and the infimum is taken over all incoherent operations $\Lambda$.
 Even more, a
closed expression for the optimal distillation rate was found recently
by Winter and Yang \cite{Winter2015}, and turns out to be equal to
the relative entropy of coherence introduced in \cite{Aaberg2006,Baumgratz2014}.  Recall the relative entropy of $\rho$ to $\sigma$ is defined as $S(\rho||\sigma)=-\mathrm{Tr}(\rho\log\sigma)-S(\rho)$, with $S(\rho)=-\mathrm{Tr}(\rho\log\rho)$ being the von Neumann entropy of $\rho$.
\begin{lem}
\label{Lem:W-Y} The distillable coherence of $\rho$ is \cite{Winter2015}
\begin{equation}
C_{d}(\rho)=C_{r}(\rho)=S(\Delta(\rho))-S(\rho),\label{eq:Cd}
\end{equation}
where $C_{r}(\rho)$ is the relative entropy of coherence, defined
as $C_{r}(\rho)=\min_{\sigma\in\mc{I}}S(\rho||\sigma)$.
\end{lem}
Note that $C_{d}(\rho)>0$ if and only if $\rho$ is not incoherent.\\

\noindent \textbf{\emph{Coherence of collaboration}}\textbf{.} We
now move to the main topic of this work, namely the assisted distillation
of coherence. As mentioned earlier, in this setting two parties Alice
and Bob share many copies of a joint state $\rho=\rho^{AB}$ and aim
to maximize coherence on Bob's system by LQICC operations.

In order to make a quantitative analysis, we define the \textit{distillable
coherence of collaboration} as the optimal rate, i.e., the optimal
number of maximally coherent states on Bob's side per copy of the
shared resource state $\rho$, in the assisted setting:
\begin{align}
C_{d}^{A|B}(\rho)=\sup\left\{ R:\lim_{n\rightarrow\infty}\left(\inf_{\Lambda}\left\Vert \Lambda\left[\rho^{\otimes n}\right]-\Phi^{\otimes\left\lfloor Rn\right\rfloor }\right\Vert \right)=0\right\} ,\label{eq:distillation}
\end{align}
where the infimum is taken over all LQICC operations
$\Lambda$. When Alice is uncorrelated from Bob, i.e. $\rho^{AB}=\rho^{A}\otimes\rho^{B}$,
then $C_{d}^{A|B}(\rho^{AB})$ reduces to the distillable coherence
$C_{d}(\rho^{B})$ which can be evaluated exactly using Lemma \ref{Lem:W-Y}
\cite{Winter2015}. In the following, we are interested in understanding
how the assistance of Alice can improve Bob's distillation rate, i.e.,
how larger $C_{d}^{A|B}(\rho^{AB})$ can be in comparison to $C_{d}(\rho^{B})$.
For answering this question, we first note that the set of bipartite
states which can be created via LQICC operations, that will be referred
to as the set $\mathcal{QI}$ of \textit{quantum-incoherent} (QI)
states, admits a simple characterization. Namely, all such states
have the following form:
\begin{equation}
\chi^{AB}=\sum_{i}p_{i}\sigma_{i}^{A}\otimes\ket{i}\bra{i}^{B}.\label{eq:QI}
\end{equation}
Here, $\sigma_{i}^{A}$ are arbitrary quantum states on $A$, and
the states $\ket{i}^{B}$ belong to the local incoherent basis of
$B$. Note that QI states have the same form as general quantum-classical
states \cite{Piani2008} (i.e., states with vanishing quantum discord
\cite{Zurek2001}), except the ``classical'' part must be diagonal
in the fixed incoherent basis.

It is obvious that any QI state has $C_{d}^{A|B}(\rho^{AB})=0$, and
the following theorem shows that the converse is true as well.

\begin{thm} \label{thm:1} A state $\rho^{AB}$ has $C_{d}^{A|B}(\rho^{AB})>0$
if and only if the state $\rho^{AB}$ is not quantum-incoherent. \end{thm}

\noindent This theorem shows that any state which cannot be created
for free via LQICC operations constitutes a resource for extracting
coherence on Bob's side. For the proof of the theorem we refer to
the Supplemental Material \cite{EPAPS}.

In the next step, we will provide an upper bound on the distillable
coherence of collaboration. For this, we introduce the QI relative
entropy:
\begin{equation}
C_{r}^{A|B}(\rho^{AB})=\min_{\chi^{AB}\in{\cal QI}}S(\rho^{AB}||\chi^{AB})\label{eq:Cr-1}
\end{equation}
with the minimization taken over the set of QI states. It is in order
to note that $C_{r}^{A|B}$ is different from the relative entropy
of discord introduced in \cite{Horodecki2005,Modi2010}, as the latter
involves a minimization over all bases of $B$, while Eq.~(\ref{eq:Cr-1})
is defined for a fixed incoherent basis $\{\ket{i}^{B}\}$. Using
the same reasoning as in \cite[see Theorem 2 there]{Modi2010}, it
is straightforward to see that $C_{r}^{A|B}$ can also be written
as
\begin{equation}
C_{r}^{A|B}(\rho^{AB})=S\left(\Delta^{B}(\rho^{AB})\right)-S(\rho^{AB})\label{eq:Cr}
\end{equation}
with $\Delta^{B}(\rho^{AB}):=\sum_{i}(\mbb{I}\otimes\op{i}{i})\rho^{AB}(\mbb{I}\otimes\op{i}{i})$.
Moreover, since the relative entropy does not increase under general
quantum operations, $C_{r}^{A|B}$ is monotonically nonincreasing
under LQICC operations. The following theorem shows that the QI relative
entropy is an upper bound on $C_{d}^{A|B}$. \begin{thm} \label{thm:2}
Given a state $\rho^{AB}$ shared by Alice and Bob, the distillable
coherence of collaboration is bounded above according to
\begin{equation}
C_{d}^{A|B}(\rho^{AB})\leq C_{r}^{A|B}(\rho^{AB}).\label{eq:bound}
\end{equation}
\par \end{thm}

\noindent The proof  can be found in \cite{EPAPS}. This result shows that in the task considered here the relative
entropy plays similar role as in the task of entanglement distillation
\cite{Horodecki2000}, bounding the distillation rate from above.
Note that for standard coherence distillation the relative entropy
of coherence is in fact equal to the optimal distillation rate \cite{Winter2015},
see also Lemma \ref{Lem:W-Y}. It is an open question if this is also
true for the task considered here, i.e., if the inequality (\ref{eq:bound})
is an equality for all quantum states $\rho^{AB}$. As we will see
in Theorem \ref{thm:3} below, at least for pure states the answer
is affirmative.\\

\noindent \textbf{\emph{Coherence of assistance}}\textbf{.} We now
introduce the \textit{coherence of assistance} (CoA) for a state $\rho$
as the maximal average coherence of the state:
\begin{equation}
C_{a}(\rho)=\max\sum_{i}q_{i}C_{r}(\psi_{i})=\max\sum_{i}q_{i}S(\Delta(\psi_{i})),\label{Eq:COA}
\end{equation}
where the maximization is taken over all pure-state decompositions
of $\rho=\sum_{i}q_{i}\op{\psi_{i}}{\psi_{i}}$, and $\psi_{i}$ is
denoting $\op{\psi_{i}}{\psi_{i}}$.

To provide CoA with an operational interpretation it is instrumental
to compare it with entanglement of assistance (EoA) originally proposed
by DiVincenzo \textit{et al.} \cite{DiVincenzo1999}. For a bipartite
state $\rho^{BC}$, one identifies a decomposition of maximal average
entanglement:
\begin{equation}
E_{a}(\rho^{BC})=\max\sum_{i}q_{i}E(\psi_{i}^{BC})=\max\sum_{i}q_{i}S(tr_{B}\psi^{BC}),\label{Eq:EOA}
\end{equation}
for $\rho^{BC}=\sum_{i}q_{i}\op{\psi_{i}}{\psi_{i}}^{BC}$. The interpretation
of EoA is that by using local measurement and one-way classical communication,
Alice can help Bob and Charlie obtain an average entanglement of at
most $E_{a}(\rho^{BC})$ when they all share $\ket{\Psi}^{ABC}$,
a purification of $\rho^{BC}$. In this case, any possible pure-state
decomposition of $\rho^{BC}$ can be realized when Alice performs
a suitable measurement and announces the result \cite{Hughston1993}.
If all the parties have access to arbitrary number of copies of the
total state $\ket{\Psi}^{ABC}$, the figure of merit is the regularized
EoA $E_{a}^{\infty}(\rho)=\lim_{n\rightarrow\infty}\frac{1}{n}E_{a}(\rho^{\otimes n})$.
For an arbitrary density matrix $\rho^{BC}$, the regularized EoA
is simply given by \cite{Smolin2005}
\begin{equation}
E_{a}^{\infty}(\rho^{BC})=\min\{S(\rho^{B}),S(\rho^{C})\}.\label{eq:EoAregularized}
\end{equation}

The CoA defined in Eq.~(\ref{Eq:COA}) has an analogous operational
meaning if we assume that the state $\rho=\rho^{B}$ belongs to Bob,
who is assisted by another party (Alice) holding a purification of
$\rho^{B}$. Through local measurement, Alice can prepare any ensemble for Bob that is compatible with $\rho^B$, which is why we take the maximization in Eq. (7).  Together with Lemma \ref{Lem:W-Y} then, $C_{a}(\rho^{B})$
quantifies a one-way coherence distillation rate for Bob when Alice
applies the same procedure for each copy of the state. In the many-copy
setting, higher one-way distillation rates can typically be obtained
when Alice performs a joint measurement across her many copies. Thus,
we consider the regularized CoA defined as $C_{a}^{\infty}(\rho):=\lim_{n\to\infty}\frac{1}{n}C_{a}(\rho^{\otimes n})$.

As we prove in \cite{EPAPS}, the CoA of a state $\rho=\sum_{i,j}\rho_{ij}\ket{i}\bra{j}$
is equal to the EoA of the corresponding maximally correlated state
\cite{Rains2001} $\rho_{\mathrm{mc}}=\sum_{i,j}\rho_{ij}\ket{ii}\bra{jj}$:
\begin{equation}
C_{a}(\rho)=E_{a}(\rho_{\mathrm{mc}}).\label{Eq:COA-EOA}
\end{equation}
Clearly, Eq.~(\ref{Eq:COA-EOA}) implies that this equality is also
true for the regularized quantities: $C_{a}^{\infty}(\rho)=E_{a}^{\infty}(\rho_{\mathrm{mc}})$.
Using Eq.~(\ref{eq:EoAregularized}), the regularized CoA thus acquires
the simple expression:
\begin{equation}
C_{a}^{\infty}(\rho)=S\left(\Delta(\rho)\right).\label{eq:CoAregularized}
\end{equation}

Equipped with these tools we are now in position to provide a closed
expression for $C_{d}^{A|B}$ for all pure states. \begin{thm} \label{thm:3}For
a pure state $\ket{\Psi}^{AB}$ shared by Alice and Bob, the following
equality holds:
\begin{equation}
C_{d}^{A|B}(\ket{\Psi}^{AB})=C_{a}^{\infty}(\rho^{B})=C_{r}^{A|B}(\ket{\Psi}^{AB})=S(\Delta(\rho^{B})).
\end{equation}
\end{thm}

\noindent The proof of the theorem can be found in \cite{EPAPS}. With Theorem \ref{thm:3} in hand, we give the von Neumann
entropy an alternative operational interpretation. Namely, let $\delta C_{d}(\rho^{B})$
denote the maximal increase in distillable coherence that Bob can
obtain when exchanging classical communication with a correlated party;
i.e. $\delta C_{d}(\rho^{B})=\max_{\rho^{AB}}[C_{d}^{A|B}(\rho^{AB})-C_{d}(\rho^{B})]$,
where the maximization is taken over all extensions $\rho^{AB}$ of
$\rho^{B}$. Noticing that the maximum is attained if $\rho^{AB}$
is pure, Lemma \ref{Lem:W-Y} and Theorem \ref{thm:3} imply that
\begin{equation}
\delta C_{d}(\rho^{B})=S(\rho^{B}).
\end{equation}
Interestingly, this result does not depend on the particular choice
of the reference incoherent basis.

Let us turn to the obvious inequality $C_{a}(\rho^{B})\leq C_{a}^{\infty}(\rho^{B})$
and ask whether $C_{a}$ is additive, in which case the inequality
becomes tight. This question is especially interesting when one considers
Ref. \cite{Winter2015} where the \textit{coherence of formation},
defined with a minimization rather than a maximization in Eq.~(\ref{Eq:COA}),
and thus a dual quantity to the CoA, is shown to be additive. Below,
we will show that in contrast, CoA fails to exhibit additivity in
general. Nevertheless, when restricting attention to $n$ copies of
an arbitrary single-qubit state $\rho$, additivity of CoA can be
proven. The latter finding is quite noteworthy since no analogous
result is known for EoA in two-qubit systems. \begin{thm} \label{Prop:COA_Qubit}
CoA is $n$-copy additive for qubit states $\rho$:
\begin{equation}
C_{a}(\rho)=C_{a}^{\infty}(\rho)=S(\Delta(\rho)).
\end{equation}
However, in general the CoA is not additive. \end{thm} We refer to
\cite{EPAPS} for the proof. It is interesting to note
that we prove non-additivity for systems with dimension 4 and above.
Thus, it remains open if $C_{a}$ is additive for qutrits. Note that
by Theorem \ref{thm:3}, this result implies that optimal coherence
distillation for single-qubit systems involves just one-way communication
and single-copy measurements from a purifying auxiliary system.\\

\noindent \textbf{\emph{Multipartite scenario}}\textbf{.} We now extend
our results to the multipartite setting. When more than one party
is providing assistance, the process of collaboratively generating
coherence for Bob's system will be called \textit{coherence localization},
in analogy to the task of \textit{entanglement localization} \cite{Verstraete2014}.

We consider $(N+1)$-partite states $\rho^{A_{1}\cdots A_{N}B}$,
where the parties $A_{1},\cdots,A_{N}$ are allowed to perform arbitrary
local quantum operations, and the party $B$ is restricted to incoherent
operations only. Additionally, classical communication is allowed
between all the parties. The aim of all the parties is to localize
as much coherence as possible on the subsystem of $B$. The corresponding
asymptotic coherence localization rate can be defined just as in Eq.~\eqref{eq:distillation}
and will be denoted by $C_{d}^{A_{1},\cdots,A_{N}|B}(\rho^{A_{1}\cdots A_{N}B})$.
For total pure states with $B$ being a qubit we find that, quite
remarkably, individual measurements on the auxiliary systems can generate
the same maximal coherence for the target system $B$ as when a global
measurement is performed across all the auxiliary systems $A_{1},\cdots,A_{N}$.
\begin{thm} Let $\ket{\Psi}^{A_{1},\cdots,A_{N}B}$ be an arbitrary
multipartite state with system $B$ being a qubit. Then
\begin{equation}
C_{d}^{A_{1},\cdots,A_{N}|B}\left(\ket{\Psi}^{A_{1},\cdots,A_{N}B}\right)=C_{d}^{A_{tot}|B}\left(\ket{\Psi}^{A_{tot}B}\right)=S\left(\Delta(\rho^{B})\right),
\end{equation}
where $A_{tot}=A_{1},\cdots,A_{N}$ is viewed as one party with the
locality constraint removed among the $A_{i}$. \end{thm} The proof
is deferred to \cite{EPAPS}. This theorem implies that
for asymptotic coherence localization the assisting parties $A_{1},\cdots,A_{N}$
do not need access to a quantum channel: local quantum operations
on their subsystems together with classical communication are enough
to ensure maximal coherence localization. This is true if the total
state is pure, and if coherence is localized on a qubit.\\

\noindent \textbf{\emph{LQICC versus SLOCC protocols}}\textbf{.} The
proof of Theorem \ref{thm:3} relied on relating the tasks of assisted
coherence distillation and assisted entanglement distillation. This
further supports a conjecture put forth in Ref.~\cite{Winter2015}
that the resource theory of coherence is equivalent to the resource
theory of entanglement for maximally correlated states \cite{Rains2001}.
We can prove a more general connection between LQICC operations in
the coherence setting and LOCC operations in the entanglement setting.

For a given bipartite state $\rho^{AB}$ we define the association
\begin{equation}
\rho^{AB}=\sum_{ij}M_{ij}^{A}\otimes\op{i}{j}^{B}\Rightarrow\tilde{\rho}^{ABC}=\sum_{ij}M_{ij}^{A}\otimes\op{ii}{jj}^{BC},\label{eq:MC}
\end{equation}
where $M_{ij}$ are operators acting on Alice's space and $\{\ket{i}\}$
is the fixed incoherent basis. As we show in \cite{EPAPS},
if two states $\rho^{AB}$ and $\sigma^{AB}$ are related via a bipartite
LQICC map, i.e. $\sigma^{AB}=\Lambda_{\mathrm{LQICC}}[\rho^{AB}]$,
then the corresponding states $\tilde{\rho}^{ABC}$ and $\tilde{\sigma}^{ABC}$
are related via a tripartite stochastic LOCC (SLOCC) map, i.e. $\tilde{\sigma}^{ABC}=\Lambda_{\mathrm{SLOCC}}[\tilde{\rho}^{ABC}]$.
Thus any procedure implementable ``for free'' in
the framework of assisted coherence has an equivalent probabilistic
``free'' implementation on the level of maximally correlated states.
We find that, in fact, for many LQICC transformations $\rho^{AB}\to\sigma^{AB}$,
the corresponding LOCC transformation $\tilde{\rho}^{ABC}\to\tilde{\sigma}^{ABC}$
can be implemented with probability one. It is an interesting open
question whether the (tripartite) LOCC analog to every (bipartite)
LQICC transformation has always a deterministic implementation.

In the case where the subsystem $A$ is uncorrelated, Eq.~(\ref{eq:MC})
reduces to $\rho=\sum_{ij}\rho_{ij}\ket{i}\bra{j}\Rightarrow\rho_{\mathrm{mc}}=\sum_{ij}\rho_{ij}\ket{ii}\bra{jj}$.
For this situation, the above results imply that for any two states
$\rho$ and $\sigma=\Lambda_{\mathrm{i}}[\rho]$ related via an incoherent
operation $\Lambda_{\mathrm{i}}$, the corresponding maximally correlated
states $\rho_{\mathrm{mc}}$ and $\sigma_{\mathrm{mc}}$ are related
via bipartite SLOCC: $\sigma_{\mathrm{mc}}=\Lambda_{\mathrm{SLOCC}}[\rho_{\mathrm{mc}}]$.
Moreover, in the asymptotic setting where many copies of $\rho$ are
available, the SLOCC procedure becomes deterministic whenever the
entanglement cost of $\sigma_{\mathrm{mc}}$ is not larger than the
distillable entanglement of $\rho_{\mathrm{mc}}$. This criterion
can be easily checked, recalling that for these states the entanglement
cost is equal to the entanglement of formation \cite{Hayden2001,Horodecki2003},
and their distillable entanglement admits a simple expression \cite{Rains2001}.
\\

\noindent \textbf{\emph{Conclusions}}\textbf{. }The results presented
above are mainly based on the new set of LQICC operations which were
introduced and studied in this work. This type of operations arises
naturally if two parties have access to a classical channel, one of
the parties can perform arbitrary quantum operations, but the other
is limited to incoherent operations only.
The results presented here can be regarded as one application of this
set of operations. Very recently, alternative applications for LQICC
were presented in \cite{Streltsov2015b,Chitambar2015}, including creation
and distillation of entanglement \cite{Chitambar2015} and implementation
of quantum teleportation in a fully incoherent manner \cite{Streltsov2015b}.
LQICC operations have also been extended to the class of local incoherent
operations (for both parties) supplemented by classical
communication \cite{Streltsov2015b,Chitambar2015}. Further applications
closely adhering to realistic physical limitations are expected in the near future.

There are in fact many scenarios of practical relevance where the task of
assisted coherence distillation can play a central
role. For instance, think of a remote or unaccessible
system on which coherence is needed as a resource (e.g. a biological
system): our results give optimal prescriptions to inject such coherence
on the remote target by acting on a controllable ancilla.
In a multipartite setting, one can imagine to distribute a correlated
state among many parties, and implement an instance of open-destination
quantum metrology, in which one party is selected to
estimate an unknown parameter \cite{Giovannetti2011} and the other
parties act locally on their subsystems in order to localize as much
coherence as possible on the chosen target, so as to enhance the estimation
precision. Similarly, the task can be a useful primitive within a
secure quantum cryptographic network \cite{GisinRMP}, in which the
distribution of non-orthogonal states (and thus coherence) is required \cite{Huelga-2013a}.

The approach presented here can also be extended to other related
scenarios. As an example, we mention the resource theory of frameness
and asymmetry \cite{Bartlett2007,Marvian2014}. The relation of these
concepts to the resource theory of coherence proposed by Baumgratz
\emph{et al}. \cite{Baumgratz2014} has been studied very recently
\cite{Marvian2015}. In this context, an important set of quantum
operations is known as thermal operations \cite{Lostaglio2015a,Lostaglio2015b}.
These operations are a subset of general incoherent operations \cite{Marvian2015}.
It will be very interesting to see how the results provided here change when local incoherent operations for one party are further restricted to local
thermal operations. This can be of direct relevance to the design of optimal ancilla-assisted work extraction protocols in thermodynamical settings
\cite{Korzekwa2015}.



\noindent \textbf{\emph{Acknowledgements}}\textbf{. }We thank Remigiusz
Augusiak for discussions. We acknowledge financial support from the
Alexander von Humboldt-Foundation, the John Templeton Foundation,
the EU IP SIQS, the EU Grant QUIC, the European Research Council (ERC
AdG OSYRIS and ERC StG GQCOP), the Spanish Ministry project FOQUS
(FIS2013-46768), the Generalitat de Catalunya project 2014 SGR 874,
the Foundational Questions Institute, DIQIP CHIST-ERA, and U.S. National
Science Foundation.

\bibliographystyle{apsrev4-1}
\bibliography{coa}

\newpage{}

\appendix*
\setcounter{equation}{0}

\section*{Supplemental Material}

\subsection{Proof of Theorem 2}

Here we will prove that any state which is not quantum-incoherent
(QI) has nonzero distillable coherence of collaboration $C_{d}^{A|B}(\rho^{AB})>0$.
To prove this, suppose that $\rho^{AB}$ is not QI. We can always
expand 
\begin{equation}
\rho^{AB}=\sum_{i,j}\ket{e_{i}}\bra{e_{j}}^{A}\otimes N_{ij}^{B},
\end{equation}
where the $\ket{e_{i}}^{A}$ form an orthonormal basis for Alice's
Hilbert space and the $N_{ij}^{B}$ are some operators on Bob's space.
Note that the operators $N_{ii}^{B}$ are nonnegative, i.e., $N_{ii}^{B}\geq0$,
and can be written as $q_{i}\rho_{i}^{B}$. The state $\rho_{i}^{B}$
can be seen as the post-measurement state of Bob if Alice performs
a von Neumann measurement in the basis $\ket{e_{i}}^{A}$, and $q_{i}$
is the corresponding probability. If for some outcome $i$ with nonzero
probability $q_{i}>0$ the corresponding state $\rho_{i}^{B}$ is
not incoherent, then Lemma 1 in the main text guarantees that $C_{d}^{A|B}(\rho^{AB})\geq q_{i}C_{r}(\rho_{i}^{B})>0$.

In the next step, we will consider the case where all the states $\rho_{i}^{B}$
corresponding to nonzero outcome probability $q_{i}>0$ are incoherent
(i.e. all the operators $N_{ii}$ are diagonal w.r.t. the incoherent
basis). Then, the condition that the state $\rho^{AB}$ is not QI
implies that $N_{kl}$ must have off-diagonal elements for some $k\neq l$.
Using the fact that $N_{kl}=N_{lk}^{\dagger}$ , we see that at least
one of the operators $N_{kl}+N_{kl}^{\dagger}$ or $i(N_{kl}-N_{kl}^{\dagger})$
must also contain offdiagonal elements in this case. Depending on
what is the case, Alice performs a von Neumann measurement in a basis
containing the state $\cos\theta\ket{e_{k}}^{A}+\sin\theta\ket{e_{l}}^{A}$
or in a basis containing the state $\cos\theta\ket{e_{k}}^{A}+i\sin\theta\ket{e_{l}}^{A}$
with some angle $\theta$ which will be determined below. In the first
case the post-measurement state of Bob $\rho_{\theta}^{B}$ is given
by 
\begin{equation}
p_{\theta}\rho_{\theta}^{B}=\cos^{2}\theta N_{kk}+\sin^{2}\theta N_{ll}+\cos\theta\sin\theta(N_{kl}+N_{lk}),\label{eq:first_case}
\end{equation}
where $p_{\theta}$ is the corresponding outcome probability. Since
$\cos^{2}\theta$, $\sin^{2}\theta$, and $\cos\theta\sin\theta$
are linearly independent, the trace of the right-hand side of Eq.~(\ref{eq:first_case})
cannot vanish for all $\theta$. Hence, there are some $0<\theta<\pi/2$
for which $p_{\theta}>0$. Similarly, in the second case the post-measurement
state of Bob $\sigma_{\theta}^{B}$ is given by 
\begin{equation}
q_{\theta}\sigma_{\theta}^{B}=\cos^{2}\theta N_{kk}+\sin^{2}\theta N_{ll}+i\cos\theta\sin\theta(N_{kl}-N_{lk})\label{eq:second_case}
\end{equation}
with outcome probability $q_{\theta}$. By the same argument, there
are some $0<\theta<\pi/2$ for which $q_{\theta}>0$. Moreover, in
both of the above cases the post-measurement state of Bob contains
offdiagonal elements.

Finally, we will now show how the above results imply that $C_{d}^{A|B}(\rho^{AB})>0$
is true for any state which is not QI. In particular, we proved that
for any such state Alice can perform a local von Neumann measurement
in such a way that the post-measurement state of Bob contains nonzero
coherence with nonvanishing probability. This means that by repeating
this procedure on each copy of $\rho^{AB}$, Bob will end up with
many copies of a state having nonzero coherence. Then, by using Lemma
1 from the main text Bob can distill maximally coherent states with
nonzero rate. This completes the proof of the theorem.

\subsection{Proof of Theorem 3}

In the following, we will prove that for any state $\rho^{AB}$ the
distillable coherence of collaboration $C_{d}^{A|B}$ is bounded above
by the QI relative entropy $C_{r}^{A|B}$: 
\begin{equation}
C_{d}^{A|B}(\rho^{AB})\leq C_{r}^{A|B}(\rho^{AB}).
\end{equation}
To prove this statement, we first note that $C_{d}$ can also be expressed
as follows: 
\begin{equation}
C_{d}^{A|B}(\rho^{AB})=\sup\left\{ C_{r}\left(\ket{\phi}\right):\lim_{n\rightarrow\infty}\left(\inf_{\Lambda}\left\Vert \Lambda\left[\rho_{i}^{\otimes n}\right]-\rho_{f}^{\otimes n}\right\Vert \right)=0\right\} ,\label{eq:Cd-1}
\end{equation}
with the initial state $\rho_{i}=\rho^{AB}\otimes\ket{0}\bra{0}^{\tilde{B}}$,
the final state $\rho_{f}=\ket{00}\bra{00}^{AB}\otimes\ket{\phi}\bra{\phi}^{\tilde{B}}$,
$\tilde{B}$ is an additional particle in Bob's hands, and the infimum
in Eq.~(\ref{eq:Cd-1}) is taken over all LQICC operations $\Lambda$
between Alice and Bob.

Then, by definition of $C_{d}^{A|B}$ in Eq.~(\ref{eq:Cd-1}), for
any $\varepsilon>0$ there exists a state $\ket{\phi}$, an integer
$n$, and an LQICC protocol $\Lambda_{n}$ acting on $n$ copies of
the state $\rho_{i}$ such that 
\begin{align}
C_{d}^{A|B}(\rho^{AB})-C_{r}(\ket{\phi}) & \leq\varepsilon,\label{eq:phi}\\
\left\Vert \Lambda_{n}\left[\rho_{i}^{\otimes n}\right]-\rho_{f}^{\otimes n}\right\Vert  & \leq\varepsilon.\label{eq:asymptotic}
\end{align}

In the next step, we will prove continuity of $C_{r}$. In particular
for two states $\rho^{XY}$ and $\sigma^{XY}$ with $||\rho^{XY}-\sigma^{XY}||\leq1$
the QI relative entropy $C_{r}$ is continuous in the following sense:
\begin{equation}
|C_{r}^{X|Y}(\rho^{XY})-C_{r}^{X|Y}(\sigma^{XY})|\leq2T\log_{2}d_{XY}+2h(T),\label{eq:continuity}
\end{equation}
where $T=||\rho^{XY}-\sigma^{XY}||/2$ is the trace distance, $d_{XY}$
is the dimension of the total system, and 
\begin{equation}
h(x)=-x\log_{2}x-(1-x)\log_{2}(1-x)
\end{equation}
is the binary entropy. It is straightforward to prove Eq.~(\ref{eq:continuity})
by using continuity of the von Neumann entropy \cite{Audenaert2007}.

The continuity relation in Eq.~(\ref{eq:continuity}) together with
Eq.~(\ref{eq:asymptotic}) implies that for any $0<\varepsilon\leq1/2$
there exists an integer $n\geq1$ and an LQICC protocol $\Lambda_{n}$
acting on $n$ copies of the state $\rho_{i}$ such that 
\begin{equation}
C_{r}^{A|B\tilde{B}}(\Lambda_{n}[\rho_{i}^{\otimes n}])\geq C_{r}^{A|B\tilde{B}}(\rho_{f}^{\otimes n})-2n\varepsilon\log_{2}d-2h(\varepsilon),
\end{equation}
where $d$ is the dimension of the total system $AB\tilde{B}$. Since
the QI relative entropy $C_{r}$ is additive and does not increase
under LQICC operations, it follows that for any $0<\varepsilon\leq1/2$
there exists an integer $n\geq1$ such that 
\begin{equation}
C_{r}^{A|B\tilde{B}}(\rho_{i})\geq C_{r}^{A|B\tilde{B}}(\rho_{f})-2\varepsilon\log_{2}d-\frac{2}{n}h(\varepsilon).
\end{equation}
By using the relations $C_{r}^{A|B\tilde{B}}(\rho_{i})=C_{r}^{A|B}(\rho^{AB})$
and $C_{r}^{A|B\tilde{B}}(\rho_{f})=C_{r}(\ket{\phi})$, the latter
inequality implies 
\begin{equation}
C_{r}^{A|B}(\rho^{AB})\geq C_{r}(\ket{\phi}).
\end{equation}
On the other hand, Eq.~(\ref{eq:phi}) means that $C_{r}(\ket{\phi})\geq C_{d}^{A|B}(\rho^{AB})-\varepsilon$.
Combining these results completes the proof of the theorem.

\subsection{Coherence of assistance and entanglement of assistance of maximally
correlated states}

In the following we will prove the relation 
\begin{equation}
C_{a}(\rho)=E_{a}(\rho_{\mathrm{mc}}),\label{eq:Ca=00003D00003DEa}
\end{equation}
where the state $\rho=\sum_{i,j}\rho_{ij}\ket{i}\bra{j}$ is arbitrary,
and the state $\rho_{\mathrm{mc}}=\sum_{i,j}\rho_{ij}\ket{ii}\bra{jj}$
is the maximally correlated state associated with $\rho$.

For proving Eq.~(\ref{eq:Ca=00003D00003DEa}), consider an optimal
decomposition of the state $\rho_{\mathrm{mc}}=\sum_{k}p_{k}\ket{\psi_{k}}\bra{\psi_{k}}$
such that 
\begin{equation}
E_{a}(\rho_{\mathrm{mc}})=\sum_{k}p_{k}E(\ket{\psi_{k}}),\label{eq:optimal-1}
\end{equation}
where the entanglement of a pure state $\ket{\psi}^{XY}$ is given
by the von Neumann entropy of the reduced state: $E(\ket{\psi}^{XY})=S(\rho^{X})$.
Note that every state $\ket{\psi_{k}}$ in the above decomposition
can be written in the form $\ket{\psi_{k}}=\sum_{i}c_{i}^{k}\ket{ii}$
with complex coefficients $c_{i}^{k}$ \cite{Horodecki2003}. In the
next step, we introduce states $\ket{\phi_{k}}=\sum_{i}c_{i}^{k}\ket{i}$,
and note that together with probabilities $p_{k}$ these states give
rise to a decomposition of the state $\rho=\sum_{k}p_{k}\ket{\phi_{k}}\bra{\phi_{k}}$.
Note that this decomposition of $\rho$ is optimal for the coherence
of assistance: 
\begin{equation}
C_{a}(\rho)=\sum_{k}p_{k}C_{r}(\ket{\phi_{k}}).
\end{equation}
The proof of Eq.~(\ref{eq:Ca=00003D00003DEa}) is complete by using
the relation $C_{r}(\ket{\phi_{k}})=E(\ket{\psi_{k}})$.

\subsection{Proof of Theorem 4}
\begin{proof}
In the following we will prove the equality 
\begin{equation}
C_{d}^{A|B}(\ket{\Psi}^{AB})=C_{a}^{\infty}(\rho^{B})=C_{r}^{A|B}(\ket{\Psi}^{AB})=S(\Delta(\rho^{B})).
\end{equation}
Clearly, the regularized CoA of a state $\rho^{B}=tr_{A}\op{\Psi}{\Psi}^{AB}$
cannot be larger than $C_{d}^{A|B}$ of its purification: 
\begin{equation}
C_{a}^{\infty}(\rho^{B})\leq C_{d}^{A|B}(\ket{\Psi}^{AB}).
\end{equation}
Together with Eq.~(11) in the main text one obtains the lower bound
\begin{equation}
S(\Delta(\rho^{B}))\leq C_{d}^{A|B}(\ket{\Psi}^{AB}).
\end{equation}
On the other hand, Eq.~(5) in the main text implies 
\begin{equation}
C_{r}^{A|B}(\ket{\Psi}^{AB})=S(\Delta(\rho^{B})).
\end{equation}
Together with Theorem 3 this completes the proof. 
\end{proof}

\subsection{Proof of Theorem 5}

In the following, we will prove the equality 
\begin{equation}
C_{a}(\rho)=C_{a}^{\infty}(\rho)=S(\Delta(\rho))\label{eq:CaQubit}
\end{equation}
for any single-qubit state $\rho$.

Let $\ket{\Psi}^{AB}$ be an arbitrary purification for $\rho^{B}$,
and expand in the incoherent basis as 
\begin{equation}
\ket{\Psi}^{AB}=\sum_{k=0}^{1}\sqrt{p_{k}}\ket{\psi_{k}}^{A}\ket{k}^{B},
\end{equation}
where $\ket{\psi_{k}}^{A}$ are arbitrary states for Alice. In the
next step we note that there always exist orthogonal states $\ket{\eta_{\pm}}^{A}$
which form a mutually unbiased basis with respect to the two states
$\ket{\psi_{k}}^{A}$. Thus, the states $\ket{\psi_{k}}^{A}$ can
be written as 
\begin{equation}
\ket{\psi_{k}}^{A}=\frac{1}{\sqrt{2}}(e^{i\alpha_{k}}\ket{\eta_{+}}^{A}+e^{i\beta_{k}}\ket{\eta_{-}}^{A})
\end{equation}
with some reals $\alpha_{k}$ and $\beta_{k}$.

When Alice performs a von Neumann measurement in the $\ket{\eta_{\pm}}^{A}$
basis, Bob will find his system in one of the post-measurement states
\begin{equation}
\ket{\phi_{\pm}}^{B}=\sqrt{p_{0}}e^{i\vartheta_{\pm}}\ket{0}^{B}+\sqrt{p_{1}}e^{i\varphi_{\pm}}\ket{1}^{B}
\end{equation}
with some reals $\vartheta_{\pm}$ and $\varphi_{\pm}$ for the +/-
outcome respectively. In both cases, the state has coherence $C_{r}(\ket{\phi_{\pm}}^{B})=S(\Delta(\rho^{B}))$.
The above reasoning shows that $C_{a}(\rho)=S(\Delta(\rho))$ is true
for any single-qubit state $\rho$. Recalling that $C_{a}^{\infty}(\rho)=S(\Delta(\rho))$
is true for any quantum state $\rho$, the proof of Eq.~(\ref{eq:CaQubit})
is complete.

We will now show that there exist states $\rho$ of dimension 4 such
that 
\begin{equation}
C_{a}(\rho)<C_{a}^{\infty}(\rho).
\end{equation}
This inequality also implies that the coherence of assistance cannot
be additive. For proving this, consider the $2\otimes4$ state 
\begin{equation}
\ket{\Psi}^{AB}=\frac{1}{2}\left(\ket{00}+\ket{11}+\ket{+2}+\ket{\hat{+}3}\right)\label{eq:Psi}
\end{equation}
with $\ket{\hat{+}}=1/\sqrt{2}(\ket{+}+i\ket{1})$. We will show that
the reduced state $\rho^{B}$ satisfies $C_{a}(\rho^{B})<C_{a}^{\infty}(\rho^{B})=2$.
We will prove this by showing a slightly stronger statement: for any
measurement of Alice performed on the state in Eq.~(\ref{eq:Psi}),
the corresponding post-measurement state of Bob will have coherence
strictly below 2.

This can be seen by contradiction: assume that for some measurement
of Alice with POVM element $M^{A}$ the corresponding post-measurement
state of Bob has maximal coherence, i.e. it corresponds to the state
$\ket{\Phi_{4}}=1/2\sum_{i=0}^{3}\ket{i}$. This condition can also
be written as follows: 
\begin{equation}
\mathrm{Tr_{A}}[M^{A}\ket{\Psi}\bra{\Psi}^{AB}]=p\ket{\Phi_{4}}\bra{\Phi_{4}}^{B},\label{eq:maximal}
\end{equation}
where $M^{A}\leq\openone^{A}$ is a nonnegative operator on the subsystem
$A$ and $p>0$ is the probability of Alice's outcome.

In the next step it is crucial to note that Eq.~(\ref{eq:maximal})
can only be fulfilled if $M$ has the same nonzero overlap with all
the states $\ket{0}$, $\ket{1}$, $\ket{+}$, and $\ket{\hat{+}}$:
\begin{equation}
\braket{0|M|0}=\braket{1|M|1}=\braket{+|M|+}=\braket{\hat{+}|M|\hat{+}}>0.
\end{equation}
Denoting the elements of $M$ by $M_{kl}=\braket{k|M|l}$, the above
equality leads to 
\begin{align}
M_{00}=M_{11} & =\frac{1}{2}\left(M_{00}+M_{11}+M_{01}+M_{10}\right)\nonumber \\
 & =\frac{1}{2}\left(M_{00}+M_{11}+iM_{01}-iM_{10}\right).
\end{align}
Taking into account that $M$ is nonnegative, this set of equations
has only one solution, namely $M_{00}=M_{11}=M_{01}=M_{10}=0$. This
completes the proof. Interestingly, from the above consideration it
is not clear if $C_{a}(\rho)$ is additive for qutrit states.

\subsection{Proof of Theorem 6}

Here we will prove the equality 
\begin{equation}
C_{d}^{A_{1},\cdots,A_{N}|B}\left(\ket{\Psi}^{A_{1},\cdots,A_{N}B}\right)=C_{d}^{A_{tot}|B}\left(\ket{\Psi}^{A_{tot}B}\right)=S\left(\Delta(\rho^{B})\right),
\end{equation}
where $B$ is a qubit, and $A_{tot}=A_{1}\cdots A_{N}$ denotes the
total system except for $B$. In the following, we assume that the
parties $A_{1},\ldots,A_{N}$ can perform arbitrary local operations,
the party $B$ is restricted to incoherent operations, and classical
communication is allowed between all parties.

For proving this statement, we will show that for some LOCC protocol
on $A_{1},\ldots,A_{N}$ all post-measurement states of $B$ will
have coherence $S\left(\Delta(\rho^{B})\right)$. This means that
by Lemma 1 of the main text the state $\ket{\Psi}^{A_{1},\cdots,A_{N}B}$
can be used to extract coherence at rate $S\left(\Delta(\rho^{B})\right)$.
This will complete the proof, since by Theorem 3 of the main text
it is not possible to achieve more coherence on $B$ even by joint
operations on $A_{1},\ldots,A_{N}$.

In the following, we will use similar arguments as in the proof of
Theorem 5. In the first step, we expand the state $\ket{\Psi}^{A_{1},\cdots,A_{N}B}$
in Bob's incoherent basis, arriving at 
\begin{equation}
\ket{\Psi}^{A_{1},\cdots,A_{N}B}=\sum_{k=0}^{1}\sqrt{p_{k}}\ket{\psi_{k}}^{A_{1},\cdots,A_{N}}\ket{k}^{B}.
\end{equation}
Similar to the proof of Theorem 5, we note that there exist orthogonal
multipartite states $\ket{\eta_{\pm}}$ which form a mutually unbiased
basis with respect to the states $\ket{\psi_{k}}$. In other words,
the states $\ket{\psi_{k}}$ can be written as 
\begin{equation}
\ket{\psi_{k}}=\frac{1}{\sqrt{2}}(e^{i\alpha_{k}}\ket{\eta_{+}}+e^{i\beta_{k}}\ket{\eta_{-}})
\end{equation}
with some reals $\alpha_{k}$ and $\beta_{k}$.

To complete the proof we will use the results of Walgate \textit{et
al.} \cite{Walgate2000}, showing that any two multipartite orthogonal
states $\ket{\eta_{+}}$ and $\ket{\eta_{-}}$ can be perfectly distinguished
via LOCC. Their results also imply the existence of a POVM $\{\Pi_{+},\Pi_{-}\}$
which can be implemented via LOCC such that 
\begin{equation}
\Pi_{+}\ket{\eta_{-}}=\Pi_{-}\ket{\eta_{+}}=0.\label{eq:discrimination}
\end{equation}
Applying this POVM on systems $A_{1}\cdots A_{N}$ of the state $\ket{\Psi}^{A_{1},\cdots,A_{N}B}$
will generate post-measurement states for Bob of the form 
\begin{equation}
\ket{\phi_{\pm}}^{B}=\sqrt{p_{0}}e^{i\vartheta_{\pm}}\ket{0}^{B}+\sqrt{p_{1}}e^{i\varphi_{\pm}}\ket{1}^{B},
\end{equation}
which leaves him with optimal coherence $C_{r}(\ket{\phi_{\pm}^{B}})=S(\Delta(\rho^{B}))$.
This completes the proof of the theorem.

\subsection{Relating LQICC and tripartite SLOCC maps}

Here we will prove that for any pair of bipartite states 
\begin{eqnarray}
\rho^{AB}=\sum_{i,j}M_{ij}^{A}\otimes\ket{i}\bra{j}^{B}, & \,\,\,\,\,\,\,\,\,\,\,\,\,\, & \sigma^{AB}=\sum_{i,j}N_{ij}^{A}\otimes\ket{i}\bra{j}^{B}\notag
\end{eqnarray}
related via $\sigma^{AB}=\Lambda_{\mathrm{LQICC}}[\rho^{AB}]$, with
an LQICC operation $\Lambda_{\mathrm{LQICC}}$, the corresponding
tripartite states 
\begin{eqnarray}
\tilde{\rho}^{ABC}=\sum_{i,j}M_{ij}^{A}\otimes\ket{ii}\bra{jj}^{BC}, & \,\,\,\,\,\,\,\,\,\,\,\,\,\, & \tilde{\sigma}^{ABC}=\sum_{i,j}N_{ij}^{A}\otimes\ket{ii}\bra{jj}^{BC}\notag
\end{eqnarray}
are related via SLOCC, i.e., $\tilde{\sigma}^{ABC}=\Lambda_{\mathrm{SLOCC}}[\tilde{\rho}^{ABC}]$
with some stochastic tripartite LOCC operation $\Lambda_{\mathrm{SLOCC}}$.
We also prove certain cases when this map can be implemented with
probability one.

Consider an LQICC protocol $\Lambda_{\mathrm{LQICC}}$ that maps $\rho^{AB}$
into $\sigma^{AB}$. In the following, we assume that this protocol
consists of $n$ intermediate LQICC operations. If we introduce the
states $\omega_{0}=\rho$ and $\omega_{n}=\sigma$, then the total
protocol can be written as $\omega_{0}^{AB}\rightarrow\omega_{1}^{AB}\rightarrow\cdots\rightarrow\omega_{n-1}^{AB}\rightarrow\omega_{n}^{AB}$.
We further suppose that each step $\omega_{k}\rightarrow\omega_{k+1}$
is either a local quantum operation on Alice's side followed by classical
communication of the outcome to Bob, or a local incoherent operation
on Bob's side, followed by classical communication of the outcome
to Alice. We will now see that for any such transformation $\omega_{k}^{AB}\rightarrow\omega_{k+1}^{AB}$
there exists a tripartite SLOCC protocol transforming $\tilde{\omega}_{k}^{ABC}$
to $\tilde{\omega}_{k+1}^{ABC}$. 

First, suppose that the process $\omega_{k}^{AB}\rightarrow\omega_{k+1}^{AB}$
involves a local measurement of Alice and classical communication
to Bob. Then, it is easy to see that the process $\tilde{\omega}_{k}^{ABC}\rightarrow\tilde{\omega}_{k+1}^{ABC}$
can be implemented deterministically, i.e., there exists a tripartite
LOCC operation such $\tilde{\omega}_{k}^{ABC}\rightarrow\tilde{\omega}_{k+1}^{ABC}$.
For this, the same local measurement has to be performed on the subsystem
$A$ of $\tilde{\omega}_{k}^{ABC}$, and the result is communicated
to both parties $B$ and $C$. 

In the following we will consider the situation where the process
$\omega_{k}^{AB}\rightarrow\omega_{k+1}^{AB}$ involves a local incoherent
operation on Bob's side, followed by classical communication to Alice.
We suppose that the state $\omega_{k}^{AB}$ has the form 
\begin{equation}
\omega_{k}^{AB}=\sum_{i,j}O_{ij}^{A}\otimes\ket{i}\bra{j}^{B}.
\end{equation}
The incoherent operation performed by Bob can always be described
by the following incoherent Kraus operators: 
\begin{equation}
K_{\alpha}^{B}=\sum_{i}c_{\alpha,i}\ket{f_{\alpha}(i)}\bra{i}^{B},\label{eq:Kraus}
\end{equation}
where $c_{\alpha,i}$ are complex numbers, and the set of functions
$f_{\alpha}(i)$ maps the set $\{i\}$ onto itself. If Bob obtains
the outcome $\alpha$, the corresponding post-measurement state is
given by 
\begin{equation}
\nu_{\alpha}^{AB}=\sum_{i,j}\frac{c_{\alpha,i}c_{\alpha,j}^{*}}{p_{\alpha}}O_{ij}^{A}\otimes\ket{f_{\alpha}(i)}\bra{f_{\alpha}(j)}^{B}
\end{equation}
with probability 
\begin{equation}
p_{\alpha}=\mathrm{Tr}\left[\sum_{i,j}c_{\alpha,i}c_{\alpha,j}^{*}O_{ij}^{A}\otimes\ket{f_{\alpha}(i)}\bra{f_{\alpha}(j)}^{B}\right].\label{eq:p}
\end{equation}
Correspondingly, the state $\tilde{\omega}_{k}^{ABC}$ has the form
\begin{equation}
\tilde{\omega}_{k}^{ABC}=\sum_{i,j}O_{ij}^{A}\otimes\ket{ii}\bra{jj}^{BC}.
\end{equation}
For showing the existence of a stochastic LOCC protocol transforming
$\tilde{\omega}_{k}^{ABC}$ to $\tilde{\omega}_{k+1}^{ABC}$ it is
enough to show that the state $\tilde{\omega}_{k}^{ABC}$ can be transformed
into the state 
\begin{equation}
\tilde{\nu}_{\alpha}^{ABC}=\sum_{i,j}\frac{c_{\alpha,i}c_{\alpha,j}^{*}}{p_{\alpha}}O_{ij}^{A}\otimes\ket{f_{\alpha}(i)}\bra{f_{\alpha}(j)}^{B}\otimes\ket{f_{\alpha}(i)}\bra{f_{\alpha}(j)}^{C}.\label{eq:nu}
\end{equation}
 via stochastic LOCC operations with nonzero probability for all $\alpha$.
This protocol consists of the following steps.
\begin{enumerate}
\item In the first step, the incoherent measurement with Kraus operators
$\{K_{\alpha}^{B}\}$ as given in Eq.~(\ref{eq:Kraus}) is performed
on the party $B$ of the total state $\tilde{\omega}_{k}^{ABC}$.
If the outcome $\alpha$ is not possible in the LQICC protocol (i.e.
if $p_{\alpha}=0$), the protocol is aborted. Otherwise, with probability
\begin{equation}
q_{\alpha}=\mathrm{Tr}[K_{\alpha}^{B}\tilde{\omega}_{k}^{ABC}(K_{\alpha}^{B})^{\dagger}]\label{eq:q}
\end{equation}
 (which is in general different from $p_{\alpha}$) the outcome $\alpha$
is obtained and broadcast to the other parties $A$ and $C$. The
corresponding post-measurement state has the form 
\begin{equation}
\tau_{\alpha}^{ABC}=\sum_{i,j}\frac{c_{\alpha,i}c_{\alpha,j}^{*}}{q_{\alpha}}O_{ij}^{A}\otimes\ket{f_{\alpha}(i)}\bra{f_{\alpha}(j)}^{B}\otimes\ket{i}\bra{j}^{C}.\label{eq:tau}
\end{equation}

\item In the next step, Charlie introduces an ancilla system $\tilde{C}$
originally in the state $\ket{0}^{\tilde{C}}$ so that the total state
is 
\begin{align}
 & \tau_{\alpha}^{ABC\tilde{C}}\\
 & =\sum_{i,j}\frac{c_{\alpha,i}c_{\alpha,j}^{*}}{q_{\alpha}}O_{ij}^{A}\otimes\ket{f_{\alpha}(i)}\bra{f_{\alpha}(j)}^{B}\otimes\ket{i}\bra{j}^{C}\otimes\ket{0}\bra{0}^{\tilde{C}}.\nonumber 
\end{align}
Depending on the outcome $\alpha$ Charlie then performs a local unitary
rotation such that 
\begin{equation}
U_{\alpha}\left(\ket{i}^{C}\ket{0}^{\tilde{C}}\right)=\ket{f_{\alpha}(i)}^{C}\ket{i}^{\tilde{C}}.\label{Eq:Bobrotate}
\end{equation}
This takes $\tau_{\alpha}^{ABC\tilde{C}}$ to the state 
\begin{align}
 & \mu_{\alpha}^{ABC\tilde{C}}\\
 & =\sum_{i,j}\frac{c_{\alpha,i}c_{\alpha,j}^{*}}{q_{\alpha}}O_{ij}^{A}\otimes\ket{f_{\alpha}(i)}\bra{f_{\alpha}(j)}^{B}\otimes\ket{f_{\alpha}(i)}\bra{f_{\alpha}(j)}^{C}\otimes\ket{i}\bra{j}^{\tilde{C}}.\nonumber 
\end{align}

\item In the final step, Charlie measures $\tilde{C}$ in the generalized
Hadamard basis: $\{\ket{b_{k}}=\frac{1}{\sqrt{d_{B}}}\sum_{j=0}^{d_{B}-1}e^{2\pi ikj/d_{B}}\ket{j}\}_{k=0}^{d_{B}-1}$.
With some probability, outcome $\ket{b_{0}}$ is obtained, leading
to the desired the final state $\tilde{\nu}_{\alpha}^{ABC}$ given
in Eq.~(\ref{eq:nu}). 
\end{enumerate}
In the following we will show that the above procedure can always
be implemented with nonzero probability. In particular, we will see
that for any $\alpha$ with probability $p_{\alpha}>0$ as described
above, the probability to obtain the state $\tilde{\nu}_{\alpha}^{ABC}$
from the state $\tilde{\omega}_{k}^{ABC}$ via tripartite SLOCC is
always nonzero. 

To prove this, we will first show that $p_{\alpha}>0$ implies $q_{\alpha}>0$,
where $q_{\alpha}$ was given in Eq.~(\ref{eq:q}). This can be seen
by contradiction, assuming that $q_{\alpha}=0$. This implies the
following:
\begin{equation}
\mathrm{Tr}\left[q_{\alpha}\tau_{\alpha}^{ABC}\openone^{AB}\otimes\ket{b_{0}}\bra{b_{0}}\right]=0,
\end{equation}
where the state $\ket{b_{0}}$ is given as $\ket{b_{0}}=\sum_{j=0}^{d_{C}-1}\ket{j}/\sqrt{d_{C}}$,
and the particles $B$ and $C$ have the same dimension. This result
together with Eq.~(\ref{eq:tau}) leads to the equality
\begin{equation}
\frac{1}{d_{C}}\mathrm{Tr}\left[\sum_{i,j}c_{\alpha,i}c_{\alpha,j}^{*}O_{ij}^{A}\otimes\ket{f_{\alpha}(i)}\bra{f_{\alpha}(j)}^{B}\right]=0.
\end{equation}
By comparing this with Eq.~(\ref{eq:p}) we see that the left-hand
side of this equality is equal to $p_{\alpha}/d_{C}$, and thus $p_{\alpha}=0$.
This proves that $p_{\alpha}>0$ implies $q_{\alpha}>0$.

To complete the proof that the above procedure can always be accomplished
with nonzero probability we note that in the measurement in the step
3 of the protocol the desired outcome appears with nonzero probability
whenever $p_{\alpha}>0$. This can be seen directly, by evaluating
the corresponding probability: 
\begin{align}
 & \mathrm{Tr}\left[\mu_{\alpha}^{ABC\tilde{C}}\openone^{ABC}\otimes\ket{b_{0}}\bra{b_{0}}\right]\nonumber \\
 & =\mathrm{Tr}\left[\sum_{i,j}\frac{c_{\alpha,i}c_{\alpha,j}^{*}}{q_{\alpha}d_{B}}O_{ij}^{A}\otimes\ket{f_{\alpha}(i)}\bra{f_{\alpha}(j)}^{B}\otimes\ket{f_{\alpha}(i)}\bra{f_{\alpha}(j)}^{C}\right].
\end{align}
By comparing this expression with Eq.~(\ref{eq:p}), we further find
that 
\begin{equation}
\mathrm{Tr}\left[\mu_{\alpha}^{ABC\tilde{C}}\openone^{ABC}\otimes\ket{b_{0}}\bra{b_{0}}\right]=\frac{p_{\alpha}}{q_{\alpha}d_{B}}.
\end{equation}
Since we assume that $p_{\alpha}>0$, this completes the proof that
the stochastic LOCC procedure discussed above has always nonzero probability
of success.

Finally, we note that for the certain types of incoherent operation
$\Lambda_{\mathrm{LQICC}}$ the aforementioned transformation is deterministic.
In particular, this is the case if the function $f_{\alpha}$ is reversible.
Then there exists a unitary rotation for Charlie $U_{\alpha}^{B}$
such that 
\begin{equation}
U_{\alpha}^{C}\ket{i}^{C}=\ket{f_{\alpha}(i)}^{C}.
\end{equation}
Performing this rotation on the state in Eq.~(\ref{eq:tau}) generates
the desired maximally correlated state $\tilde{\nu}_{\alpha}^{ABC}$,
and steps 2 and 3 in the above protocol are omitted.

In summary, the transformation $\tilde{\rho}^{ABC}\to\tilde{\sigma}^{ABC}$
can always be achieved with some nonzero probability. If all the incoherent
operations in $\Lambda_{\mathrm{LQICC}}$ have Kraus operators $K_{\alpha}$
with $f_{\alpha}(i)$ being reversible for every $\alpha$, then the
transformation can be accomplished with probability one.
\end{document}